\documentclass[11pt,dvips]{article}
\usepackage[dvips]{graphicx}
\usepackage{color}
\usepackage{amsmath,amssymb,amsfonts}

\newcommand{\s}{\,{\rm s}}
\newcommand{\GeV}{\,{\rm GeV}}
\newcommand{\keV}{\,{\rm keV}}
\newcommand{\eV}{\,{\rm eV}}
\newcommand{\Mpc}{\,{\rm Mpc}}
\newcommand{\fex}{{\it e.g.} }
\newcommand{\comp}{{\it cf.} }
\newcommand{\ppp}[1]{\begin{pmatrix} #1 
\end{pmatrix}}

\title{\vspace{-1.5cm}
{\normalsize\rightline{DESY 08-128}\rightline{TUM-HEP 699/08}}\ \vskip 1cm
\bf\boldmath 
Hidden gauginos of an unbroken $U(1)$: Cosmological constraints and phenomenological prospects
\vspace{11mm}} 

\author{A.~Ibarra$^{a,b}$, A.~Ringwald$^{b}$, C.~Weniger$^{b}$\\[4mm]
{\normalsize\it a  Physik Department T30, Technische Universit\"at M\"unchen,}\\[-0.05cm]
{\it\normalsize James-Franck-Strasse, 85748 Garching, Germany.}\\[2mm]
{\normalsize\it b Deutsches Elektronen-Synchrotron DESY,}\\
{\it\normalsize Notkestrasse 85, 22607 Hamburg, Germany.}\\[2mm]
\date{\empty}
}

\begin{document}
\begin{titlepage} 
  \maketitle
\begin{abstract}
  We study supersymmetric scenarios where the dark matter is the gaugino of an
  unbroken hidden $U(1)$ which interacts with the visible world only via a
  small kinetic mixing with the hypercharge.  Strong constraints on the
  parameter space can be derived from avoiding overclosure of the Universe and
  from requiring successful Big Bang Nucleosynthesis and structure formation.
  We find that for typical values of the mixing parameter, scenarios with
  neutralino NLSP are excluded, while scenarios with slepton NLSP are allowed
  when the mixing parameter lies in the range $\chi\sim\mathcal{O}
  (10^{-13}-10^{-10})$. We also show that if the gravitino is the LSP and the
  hidden $U(1)$ gaugino the NLSP, the bounds on the reheating temperature from
  long lived charged MSSM relics can be considerably relaxed and we comment on
  the signatures of these scenarios at future colliders.  Finally, we discuss
  the case of an anomalously small mixing, $\chi\ll 10^{-16}$, where the
  neutralino becomes a decaying dark matter candidate, and derive constraints
  from gamma ray experiments.
\end{abstract}

\thispagestyle{empty}
\end{titlepage}
\newpage \setcounter{page}{2}

\section{Introduction}
The existence of dark matter in the Universe is perhaps the most solid
indication for physics beyond the Standard Model of particle
physics~\cite{BHS05}. Among the many extensions of the Standard Model that have
been proposed in recent years, supersymmetry (SUSY) arguably remains as the
most popular.  Among other merits, supersymmetric scenarios provide a very
promising candidate for the dark matter: the lightest supersymmetric particle
(LSP)~\cite{EHN+84,JKG96}. With the particle content of the Minimal
Supersymmetric Standard Model (MSSM), the LSP can be, over a large range of
parameters, either the lightest neutralino, the lightest sneutrino or the
lightest stau. Among these, only the lightest neutralino is still allowed by
present experiments as a viable dark matter candidate, provided $R$-parity is
almost exactly conserved. Furthermore, if supersymmetry is promoted to a local
symmetry, the particle content of the MSSM also includes the gravitational
supermultiplet, of which the spin $3/2$ component, the gravitino, is also a
viable dark matter candidate~\cite{PP82}, even if $R$-parity is slightly
violated~\cite{TY00,BCH+07}.

On the other hand, many extensions of the MSSM contemplate the possibility of a
hidden sector, consisting of superfields which are singlets under the Standard
Model gauge group.  Hidden sector superfields usually couple very weakly to our
observable sector, thus constituting a very natural arena for finding dark
matter candidates.  Generically, hidden sector particles couple to our
observable sector only via non-renormalizable operators, presumably suppressed
by a large mass scale, with a structure that is strongly model dependent.  In
consequence, deriving implications of the hidden sector dark matter on the
thermal history of the observable Universe and for future collider experiments
is hindered by our complete ignorance of the strength and the structure of the
hidden sector interactions with our observable sector. 

There are however three instances where the hidden sector particles can couple
to the MSSM particles via renormalizable operators, with a structure which is
well defined by the Lorentz and gauge symmetries. Firstly, a hidden sector
chiral superfield, $S$, could couple to the lepton and up-type Higgs doublet
superfields via the Yukawa coupling $S H_u L$ in the superpotential, or to the
two MSSM Higgs doublets, via $S H_u H_d$, provided these terms are also
invariant under the hidden sector gauge group (as well as possible discrete and
global symmetries of the theory).  Secondly, if one of the
MSSM chiral superfields is charged under a hidden sector gauge group, it will
interact with the corresponding hidden sector gauge superfield, and in turn
with other hidden sector chiral superfields via the $D$-term.  Finally, a
hidden sector abelian vector superfield, $X$, may couple to the hypercharge
vector superfield through a kinetic mixing term, which is always allowed by the
gauge symmetries \cite{Holdom86, FH91}.

In this paper we will concentrate on the last situation and we will study
scenarios where the hidden $U(1)$ gauge group is unbroken at low energies. If
this is the case, the corresponding hidden $U(1)$ gaugino will have a mass
comparable to the typical soft SUSY breaking masses of the observable sector
particles and, in some instances, smaller. Therefore, the hidden $U(1)$ gaugino
will not only be the lightest supersymmetric particle of the hidden sector, but
also the lightest among all the supersymmetric particles.

Our motivation to consider an unbroken hidden sector $U(1)$ group is twofold.
First, in string theory compactifications hidden sector $U(1)$ groups are
ubiquitous and some of them could remain unbroken at low energies, in complete
analogy to the familiar electromagnetic $U(1)$ of our observable sector.
Secondly, the case of the hidden sector unbroken $U(1)$ is particularly
intriguing, since this situation is practically unconstrained by present
experiments. Indeed, it was shown long ago by Holdom that, in a
non-supersymmetric world, the hidden $U(1)$ gauge boson (the ``paraphoton''
\cite{Okun82}) completely decouples from the observable sector \cite{Holdom91}.
This result can be also generalized to a supersymmetric theory. Let us consider
the SUSY invariant part of the Lagrangian,
\begin{eqnarray}
  \nonumber
  {\cal L}&=&\int d^2\theta \, \left(\hat{W}_B^\alpha \hat{W}_{B\,\alpha}+
  \hat{W}_X^\alpha \hat{W}_{X\,\alpha}+2\chi \hat{W}_B^\alpha
  \hat{W}_{X\,\alpha}\right)+ \text{h.c.}+\\ 
  &&+\int d^2\theta d^2\bar \theta \, \left(\Phi^\dagger e^{Y g_Y \hat{B}}
  \Phi+ h^\dagger e^{q g_X \hat{X}} h\right)\;,
  \label{eqn:Lagrangian1}
\end{eqnarray}
where the field strength superfield is defined as
$\hat{W}_V^\alpha=-\frac{1}{4} {\bar D}{\bar D} D^\alpha \hat{V}$,
$\hat{V}=\hat{B}, \hat{X}$ being the hypercharge or the hidden $U(1)$ vector
superfield, while $\Phi$ and $h$ denote, respectively, any Standard Model or
hidden sector chiral superfield. Finally, $\chi$ is the kinetic mixing
parameter, which is induced through quantum effects by chiral superfields
charged under both gauge groups.  Without additional symmetries, values around
$\chi\sim 10^{-3} - 10^{-4}$ are naturally obtained.  However, \fex in
compactifications of heterotic \cite{DKM97, BMW06} and type II \cite{LS07,
AS04, BHK05, AJK+08, AGJ+08} strings, much smaller mixings are possible.  A
lower bound around $\chi \gtrsim 10^{-16}$ was argued to hold in cases of gauge
mediated supersymmetry breaking in heterotic string models \cite{DKM97},
whereas in type II models with warped extra dimensions the kinetic mixing
parameter can be exponentially small \cite{AGJ+08}. 

The gauge kinetic terms in Eq.~\eqref{eqn:Lagrangian1} can be made canonical by
introducing shifted vector superfields,
\begin{eqnarray}
  X&=& \hat{X}+ \chi \hat{B}\;, \\
  B&=& \sqrt{1- \chi^2} \hat{B}\;,
\end{eqnarray}
leading to
\begin{eqnarray}
  {\cal L}&=&\int d^2\theta \, \left(W_B^\alpha W_{B\,\alpha}+
  W_X^\alpha W_{X\,\alpha}\right) + \text{h.c.}+ \\\nonumber
  &&+\int d^2\theta d^2\bar \theta \, \left(\Phi^\dagger 
  e^{Y g^\prime_Y B} \Phi+
  h^\dagger e^{q g_X X- q g^\prime_X B} h\right)\;,
\end{eqnarray}
where $g^\prime_Y= g_Y/\sqrt{1-\chi^2}$ and $g^\prime_X= \chi
g_X/\sqrt{1-\chi^2}$.  Therefore, the canonical normalization of the kinetic
terms produces an unobservable shift of the hypercharge gauge coupling and the
generation of a ``minihypercharge'' for the hidden sector chiral superfields
\cite{Holdom86}.  Different astrophysical observations and laboratory
experiments constrain the possible values of the minihypercharge and the masses
of the hidden sector particles. For instance, one obtains $\chi \lesssim
10^{-13}$ for masses below $10\keV$ from plasmon decay in red giants
\cite{DP94, DHR00} (see also Ref.~\cite{MPS07}). Nevertheless these bounds can
be automatically avoided if the masses are large. In this case, as long as
supersymmetry remains unbroken, the hypercharge vector superfield completely
decouples from the observable sector and is not subject to any experimental
constraint. 

The breaking of supersymmetry changes dramatically the previous picture.
Although the hidden $U(1)$ gauge boson remains decoupled from the observable
sector, we will show that  in the presence of SUSY breaking effects a mixing
between the hidden $U(1)$ gaugino and the MSSM neutralinos is induced.  Then,
the unbroken hidden $U(1)$ might produce observable effects in the cosmological
evolution of the Universe, at collider experiments or in cosmic ray fluxes.

Models with a hidden $U(1)$ extension of the SM or MSSM have been extensively
studied in the literature (see \fex \cite{CNW06}, for a recent review see
Ref.~\cite{Langacker08}). Some of these works take into account kinetic- and
mass-mixing \cite{FKN07, FLN07, KN05, KW06, KN04, CY07, PRV08a}, but it is
typically assumed that the gauge symmetry of the additional $U(1)$ is broken by
a Higgs or Stueckelberg mechanism.\footnote{Exceptions are \fex
Ref.~\cite{Dobrescu05, BBS08}, where constraints on, and consequences of,
higher dimensional operators that couple hidden and observable sector are
studied. In Ref.~\cite{FTY08} the authors study BBN and CMB constraints on the
particle content of a completely decoupled hidden sector which may contain
unbroken $U(1)$s.  Furthermore, see Ref.~\cite{Redondo08a} for a short
discussion about gauge coupling unification in the presence of kinetic mixing.}
In this case, it is possible to derive bounds from high precision LEPI data if
the hidden $U(1)$ gauge boson mass is large (namely, $\chi\lesssim 0.05$ for
masses around $200\GeV$ \cite{FLN07}), and from different astronomical
observations and laboratory experiments if the masses are small (namely, for
masses around $100\eV$, the observed lifetime of the sun translates into a
bound $\chi\lesssim 10^{-13}$ \cite{Redondo08}, see Ref.~\cite{JRR08} for
bounds from considerations of the hidden CMB).

The purpose of this paper is to consider in some detail cosmological
constraints and phenomenological properties of a kinetically mixed hidden
$U(1)$ extension of the MSSM with unbroken gauge symmetry. We will assume that
the hidden $U(1)$ gaugino is the LSP in most of the paper, and concentrate on
its prospect of being dark matter.  In section 2, we will describe the model in
the component formalism and discuss typical values for mass- and
mixing-parameters in scenarios with gauge and gravity mediation of
supersymmetry breaking.  In section 3, we derive a bound on the mixing
parameter from thermal overproduction of the hidden $U(1)$ gaugino. Section 4
considers bounds from primordial nucleosynthesis for the cases of stau and
neutralino NLSPs. There, we also analyze the effects of a possibly light
gravitino and show how bounds on the reheating temperature can be relaxed.  In
section 5 we briefly discuss the collider phenomenology of this scenario.  In
section 6 we study the case with an anomalously small kinetic mixing, where the
neutralino NLSP could become a viable, though unstable, dark matter candidate.
Finally, we present our conclusions and an outlook in section 7.

\section{Model}
\label{sec:Model}
In this section, we will briefly describe the model under consideration in the
more familiar component formalism. In the Wess-Zumino gauge, a vector
superfield can be expanded in component fields as $V=-\theta \sigma^\mu {\bar
\theta} V_\mu +i \theta\theta {\bar \theta} {\bar \lambda}-i {\bar \theta}
{\bar \theta}\theta  \lambda+ \frac{1}{2} \theta\theta {\bar \theta}{\bar
\theta} D$. Then, the pure gauge part of Eq.~\eqref{eqn:Lagrangian1}, including
supersymmetry breaking effects, reads:
\begin{eqnarray}
  \label{eqn:Lagrangian}
  \mathcal{L}_{gauge}&
  =&-\frac{1}{4} \ppp{\hat{X}_{\mu\nu}&\hat{B}_{\mu\nu}} \mathcal{K} 
  \ppp{\hat{X}^{\mu\nu}\\ \hat{B}^{\mu\nu}}
  -i\ppp{ \hat{\lambda}_X^\dagger & \hat{\lambda}_B^\dagger}\mathcal{K}
  \bar{\sigma}^\mu\partial_\mu\ppp{\hat{\lambda}_{X}\\\hat{\lambda}_{B}}
  \\\nonumber&& 
  +\frac{1}{2}\ppp{\hat{D}^*_X &
  \hat{D}^*_B}\mathcal{K}\ppp{\hat{D}_{X}\\\hat{D}_{B}}
  -\frac{1}{2}\ppp{\hat{\lambda}_X &
  \hat{\lambda}_B}\hat{\mathcal{M}}
  \ppp{\hat{\lambda}_{X}\\\hat{\lambda}_{B}} + \text{h.c.}\;,
\end{eqnarray}
where $\mathcal{K}$ and $\hat{\mathcal{M}}$ denote, respectively, the kinetic
and mass mixing matrices:
\begin{eqnarray}
  \mathcal{K}=\ppp{1&\chi\\\chi&1}\hspace{0.5cm}\text{and}\hspace{0.5cm}
  \hat{\mathcal{M}}=\ppp{\hat{M}_X&\delta\hat{M}\\\delta\hat{M}&\hat{M}_B}\;.
  \label{eqn:MixingMatrix}
\end{eqnarray}

It is convenient to work in the basis where the kinetic terms are canonical.
To this end, we will redefine the vector superfields according to
\begin{eqnarray}
  \ppp{\hat{X}\\\hat{B}}=
  \ppp{1&-\frac{\chi}{\sqrt{1-\chi^2}}\\0&\frac{1}{\sqrt{1-\chi^2}}}\ppp{X\\B}\;.
  \label{eqn:GL2Transformation}
\end{eqnarray}
As explained in the introduction, the canonically normalized fields
$X_{\mu\nu}$ and $D_X$ completely decouple from the observable sector. On the
other hand, a mixing between the hidden gaugino and the bino remains, unless
$\delta\hat{M} \simeq \chi \hat{M}_X$.  More precisely, in the basis where the
kinetic terms are canonical, the extended neutralino $(5\times5)$ mass matrix
reads, to lowest order in $\chi$, 
\begin{eqnarray}
  \mathcal{M}_\text{N}=\ppp{ M_X & \delta M & 0 & 0 & 0 \\ 
  \delta M & M_B & 0 & -M_Z c_\beta s_W & M_Z s_\beta s_W \\
  0 & 0 & M_W & M_Z c_\beta c_W & -M_Z s_\beta c_W \\
  0 & -M_Z c_\beta s_W & M_Z c_\beta c_W & 0 & -\mu \\
  0 & M_Z s_\beta s_W & -M_Z s_\beta c_W & -\mu & 0 }\;,
  \label{eqn:NeutralinoMassMatrix}
\end{eqnarray}
where, $\delta M \simeq \delta\hat{M} -\chi \hat{M}_X$, $M_X\simeq\hat{M}_X$
and $M_B\simeq\hat{M}_B$.  Here, $\mu$ denotes the MSSM $\mu$-term, $M_Z$ the
mass of the $Z^0$ gauge boson, $s_W$ the sine of the Weinberg angle and
$s_\beta$ is related to the ratio of the two Higgs VEVs.

Lastly, as was mentioned in the introduction,  hidden sector matter particles
charged under the hidden $U(1)$ will acquire a minihypercharge, as a
consequence of the canonical normalization of the kinetic terms. For simplicity
we will assume throughout the paper that hidden sector matter particles are too
heavy to be produced during reheating, and thus do not influence the evolution
of the Universe.

The size of the gaugino mass terms $\hat {\cal M}$ in the Lagrangian,
Eq.(\ref{eqn:Lagrangian}), is very model dependent.  For example, in models
with gravity mediation, gaugino masses would arise from the term
\begin{eqnarray}
  \mathcal{L}\supset\int d\theta^2 d{\bar \theta}^2 \frac{1}{M_P}
  (\hat{Z}^\dagger_1 \hat{W}_B \hat{W}_B +
  2 \chi \hat{Z}^\dagger_2  \hat{W}_B \hat{W}_X + \hat{Z}^\dagger_3 
  \hat{W}_X \hat{W}_X). + {\rm h.c.}\;,
  \label{eqn:LagrangianSuperFields}
\end{eqnarray}
where coefficients of order one have been dropped. In this equation $\hat Z_i$
are three spurion superfields, which will eventually acquire a vacuum
expectation value, $\langle \hat{Z}_i \rangle= M_P+\theta^2 F_i$, by some
unspecified mechanism\footnote{We assume that breaking of supersymmetry takes
place in a sector that is distinct from the hidden $U(1)$ under consideration,
hence the $D$-terms are zero and kinetic mixing has no effect on the MSSM mass
spectrum (like \fex in Ref.~\cite{DKM97}).}, yielding $\hat M_B=F_1/M_P$,
$\delta \hat M =\chi F_2/M_P$, $\hat M_X=F_3/M_P$, where $M_P\simeq
2.4\times10^{18}\GeV$ denotes the reduced Planck mass. 

The simplest case consists on assuming just one spurion superfield which
couples universally to all vector superfields, namely $Z_i\equiv Z$ for all
$i$.  If this is the case, at the scale $M_P$, the different soft terms satisfy
the relations $\hat{M}_B=\hat{M}_X=\sqrt{3} m_{3/2}$ and
$\delta\hat{M}=\sqrt{3} \chi m_{3/2}$, $m_{3/2}$ being the gravitino mass.
Therefore, after the canonical normalization of the gaugino kinetic terms, the
mass mixing term  $\delta M $ in Eq.~\eqref{eqn:NeutralinoMassMatrix} vanishes.
However, the universality of the gaugino mass terms is broken by quantum
effects, thus inducing a non-vanishing $\delta M $ at low energies.  Using the
renormalization group equations for the gaugino mass matrix in
appendix~\ref{app:RG}, we find that the mass mixing at the electroweak scale
reads
\begin{eqnarray}
  \delta M_\text{EW}\simeq\frac{1}{8\pi^2}g_X^2 B_{XX} \ln 
  \left( \frac{M_P}{M_\text{hid}} \right) \chi_\text{EW} M_X\;,
  \label{eqn:MassMixing}
\end{eqnarray}
where $g_X$ is the gauge coupling of the hidden $U(1)$ gauge group,
$M_\text{hid}$ is the mass scale of hidden sector matter, and $B_{XX}=\sum_i
Q^2_i$ is the sum, for all the particles in the theory, of all the hidden
$U(1)$ charges squared.  On the other hand, the renormalization group evolution
will also make the hidden gaugino mass smaller at low energies.  If there is an
enough number of (heavy) matter states in the hidden sector, the hidden gaugino
will become the lightest supersymmetric particle. 

In more general scenarios with gravity mediation, where several spurion fields
contribute to the breaking of supersymmetry, a tree level mixing between the
bino and the hidden gaugino will usually remain after canonical normalization
of the kinetic terms. On the other hand, the nature of the lightest
supersymmetric particle, whether it is the gravitino, the bino or the hidden
gaugino, depends on the details of the model. 

The most natural scenarios with a light hidden $U(1)$ gaugino are probably
those with gauge mediated supersymmetry breaking. Assuming that the messenger
fields are not charged under the hidden $U(1)$ gauge group, the soft mass of
the hidden gaugino will entirely come from gravity effects.  If this is the
case, the gravitino and the hidden $U(1)$ gaugino will acquire similar masses,
which are naturally much smaller than the masses of the supersymmetric
particles of the MSSM. Again, whether the LSP is the gravitino or the hidden
gaugino depends on the particle content of the hidden sector, which will drive
the hidden gaugino mass to smaller values at low energies through radiative
effects. 

An important quantity in our analysis will be the mixing angle $\Theta$ between
the bino and the hidden $U(1)$ gaugino mass and interaction
eigenstates\footnote{Note that we always assume the mass hierarchy $M_B < M_W
<\mu$ in this paper.}. The approximate diagonalization of
Eq.~\eqref{eqn:NeutralinoMassMatrix} yields
\begin{eqnarray}
  \Theta \simeq \frac{\delta M^\text{EW}}{M_B^\text{EW}-M_X^\text{EW}}\;,
  \label{eqn:ChiEffective2}
\end{eqnarray}
where we have emphasized that all the quantities should be evaluated at the
electroweak scale. In the scenarios of supersymmetry breaking proposed in this
section with hidden $U(1)$ gaugino as LSP, the mixing angle $\Theta$ can be
written as
\begin{eqnarray}
  \Theta \simeq C \cdot \chi_\text{EW}
  \frac{M_X^\text{EW}}{M_B^\text{EW}-M_X^\text{EW}}\;,
  \label{eqn:ChiEffective}
\end{eqnarray}
where $C$ is a constant that can roughly vary between $10^{-2} \lesssim C
\lesssim 1$, depending on whether $\delta M^\text{EW}$ is already present at
tree level or is generated radiatively, and on the particle content of the
hidden sector. For definiteness we will choose $C=1$ in the subsequent
analysis. 

Following the above discussion, we will first study the phenomenology of a
hidden $U(1)$ gaugino LSP, and second the case where both, the hidden $U(1)$
gaugino and the gravitino, constitute the lightest supersymmetric particles.

\section{Overproduction constraints}
\label{sec:OP}
The least model dependent bound on the hidden $U(1)$ gaugino scenario comes
from considerations of the thermal history of the universe.  After reheating of
the visible sector, hidden $U(1)$ gauginos can be produced in the hot primeval
plasma through the mixing with the MSSM neutralinos. 

This mechanism is similar to the thermal production of gravitinos or axions
(see \fex Ref.~\cite{BBB01,PS07,RS07}).  However, in contrast to these cases,
the final abundance of the hidden $U(1)$ gaugino does not depend on the
reheating temperature, $T_R$. This follows from the fact that mass- and
kinetic-mixings appear at the renormalizable level of the Lagrangian, whereas
gravitino and axion couplings are mass-suppressed. Actually, the production is
most efficient at low temperatures, and this makes an exact calculation of the
hidden $U(1)$ gaugino abundance extremely complicated, since at temperatures
around $T\sim100\GeV$ the QCD coupling constant $g_s$ is of the order of one,
which precludes a sharp separation between hard, $T$, and soft, $ g_s T$,
momenta as required for the proper treatment of  hard thermal loops and
screening effects~\cite{BY91}. How to deal with this drawback is an open
problem even in the milder case of gravitino production (see, however,
Ref.~\cite{RS07}). 

For the purposes of this paper it is enough to derive an order of magnitude
estimate on the hidden $U(1)$ gaugino abundance. To this end, we have
calculated the collision integral in a relativistic QCD plasma, taking into
account $2\rightarrow2$ scattering in the QCD and hypercharge sector with a
bino in the final state.  Since binos mix with hidden $U(1)$ gauginos with the
mixing angle $\Theta$, the production rate of hidden $U(1)$ gauginos can be
straightforwardly calculated by multiplying the result for the binos by
$\Theta^2$.  The relevant scattering processes and the corresponding squared
matrix elements are listed in Tab.~\ref{tab:MatrixElements}.
\begin{table}
  \centering
  \begin{tabular}{|l|c|}
    \hline
    Process & $\overline{|\mathcal{M}|^2} / \Theta^2 (g' g_s)^2 |T_{ba}^j|^2 Y_{L,R}^2$ \\\hline\hline
    1) $qg\rightarrow \tilde{q}\lambda_X$ & $-2\frac{u}{s}$ \\\hline
    2) $\tilde{q}\tilde{g}\rightarrow \tilde{q}\lambda_X$ & $-2\left( \frac{t}{s}+\frac{s}{t} \right)$ \\\hline
    3) $\tilde{q}g\rightarrow q\lambda_X$ & $2\frac{u}{t}$ \\\hline
    4) $q\tilde{g}\rightarrow q\lambda_X$ & $4$ \\\hline
    5) $q\bar{\tilde{q}}\rightarrow g\lambda_X$ & $-4\frac{s}{u}$ \\\hline
    6) $q\bar{q}\rightarrow \tilde{g}\lambda_X$ & $8$ \\\hline
    7) $\tilde{q}\bar{\tilde{q}}\rightarrow \tilde{g}\lambda_X$ & $4\left( \frac{t}{u}+\frac{u}{t} \right)$ \\\hline
  \end{tabular}
  \caption{Squared matrix elements for inelastic two-body scatterings with a
  hidden $U(1)$ gaugino in the final state.  Here, $T_{ba}^i$ denotes the
  generators of the fundamental representation of the $SU(3)$ gauge group,
  $Y_{L,R}$ the hypercharges of the (s)quarks, and $g'$ and $g_s$ the gauge
  couplings of the electroweak and strong interactions, respectively. We
  average over initial and sum over final spins.}
  \label{tab:MatrixElements}
\end{table}

Summing over all spins and particles, the collision term of hidden $U(1)$
gauginos in the relativistic MSSM plasma reads\footnote{We use Boltzmann
distribution functions for simplicity.}
\begin{eqnarray}
  \gamma_\text{QCD}\equiv\frac{d^4n_X}{dVdt}&=&
  \frac{308}{3\pi^3}\alpha' \alpha_s  \Theta(T)^2
  \left( 1-\frac{4}{7}\gamma_E-\frac{4}{7}\ln\frac{k^\ast}{T} \right)T^4 
  \nonumber\\
  &\simeq& 3\times 10^{-3} \Theta(T)^2  T^4 \;,
  \label{eqn:PrductionRate}
\end{eqnarray}
where one has in general to take into account the temperature dependence of the
mixing angle $\Theta$, which stems from the thermal mass of the bino (see
Eq.~\eqref{eqn:ChiEffective}). Here, we simply take $M_B(T)\simeq M_B$, since
most of the production is expected to happen when the particles become
non-relativistic. In this equation, $\alpha'$ and $\alpha_s$ denote the
hypercharge and QCD couplings, respectively, and $\gamma_E\simeq0.577$.
Furthermore, $k^\ast$ denotes the cut-off of intermediate three-momenta in
$t$-channel processes. Only processes with quarks in the intermediate state
contribute to these divergences, and we identify $k^\ast$ with their thermal
mass (at $T\sim 100\GeV$ the thermal quark masses lie between $0.63T$ and
$0.84T$ \cite{RS07}).  Finally, solving the Boltzmann equation yields the
thermal abundance of hidden $U(1)$ gauginos,
\begin{eqnarray}
  \Omega_X h^2 \approx 5.5\times 10^7\left( \frac{M_X}{100\GeV} \right) 
  \int_{T_0}^{T_R} dT
  \frac{M_P}{T^2}\frac{\gamma_\text{QCD}(T)\Theta^2(T)}{T^4}\;.
  \label{eqn:ThermalAbundance}
\end{eqnarray}
Taking into account thermal masses could actually lead to a further order-one
enhancement due to particle decay in the QCD plasma \cite{RS07}.  Contributions
from electroweak interactions and Yukawa couplings are also expected to give
sizable corrections, again of order one.

Requiring that $\Omega_X h^2 \lesssim 0.1$ gives the overproduction bound
\begin{eqnarray}
  \Theta\lesssim 5\times10^{-12}\left( \frac{M_X}{M_B} \right)^{-1/2} \;,
  \label{eqn:OverProductionBound}
\end{eqnarray}
where we have used $T_0\simeq M_B$ as a cutoff in
Eq.~\eqref{eqn:ThermalAbundance}, which corresponds to squark- and gluino
masses around $3M_B$. For other values the bound scales roughly like
$\sqrt{T_0}$.

\begin{figure}[h]
  \begin{center}
    \includegraphics[width=0.9\linewidth]{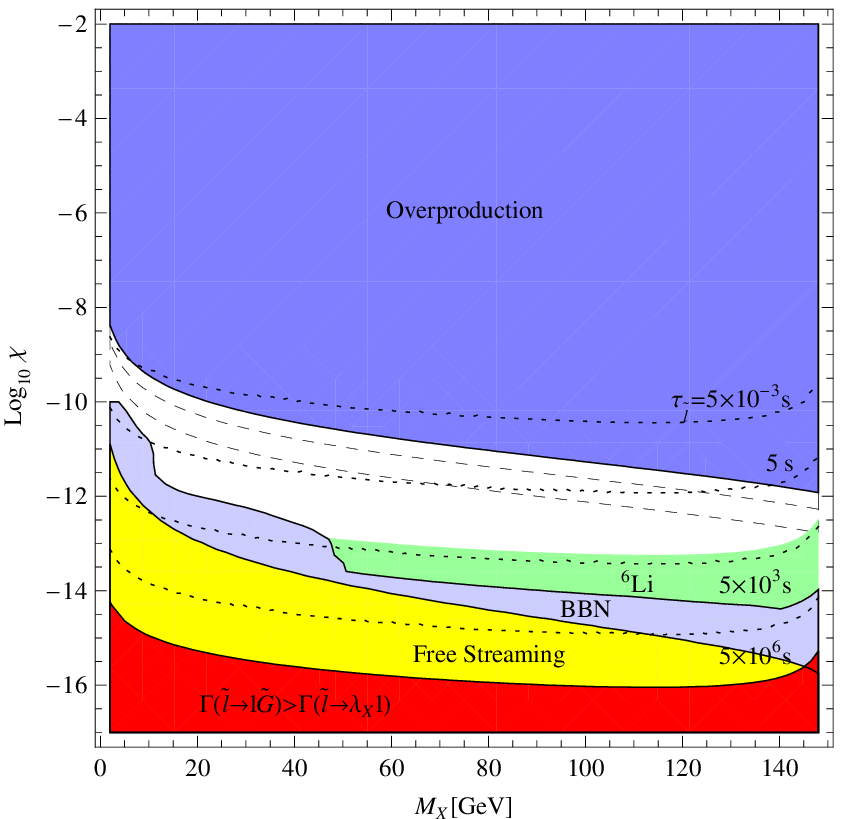}
    \vspace{-0.5cm}
  \end{center}
  \caption{Summary of bounds on the hidden $U(1)$ gaugino parameter space for
  the case of a slepton NLSP. We use $M_{\tilde{l}}=150\GeV$ and $M_B=180\GeV$.
  The upper dark blue region is excluded by thermal overproduction. Below this
  region, the hidden $U(1)$ gaugino is dominantly produced via late decaying
  sleptons. The light blue region is excluded by energy injection during BBN
  \cite{KKM05}, whereas the light green region is excluded by catalysis of
  $^6\text{Li}$ production \cite{Pospelov07}. We also show the region which
  would be excluded solely by free streaming arguments. The dotted lines show
  the slepton lifetime. In the presence of a gravitino with $M_{\tilde{G}} =
  100\GeV$ the slepton would dominantly decay into the hidden $U(1)$ gaugino,
  except in the red lower region. The dashed lines show the region that is
  potentially excluded by bounds on $\Lambda$CWDM models in scenarios where the
  thermally produced hidden $U(1)$ gaugino decays into a gravitino LSP with
  large free streaming length (see Fig.~\ref{fig:FSXandGravitino}). The lines
  correspond to a dark matter fraction $f=0.02$ and $f=0.2$ which is allowed to
  be warm.}
  \label{fig:boundsSlepton}
\end{figure}

\begin{figure}[h]
  \begin{center}
    \includegraphics[width=0.9\linewidth]{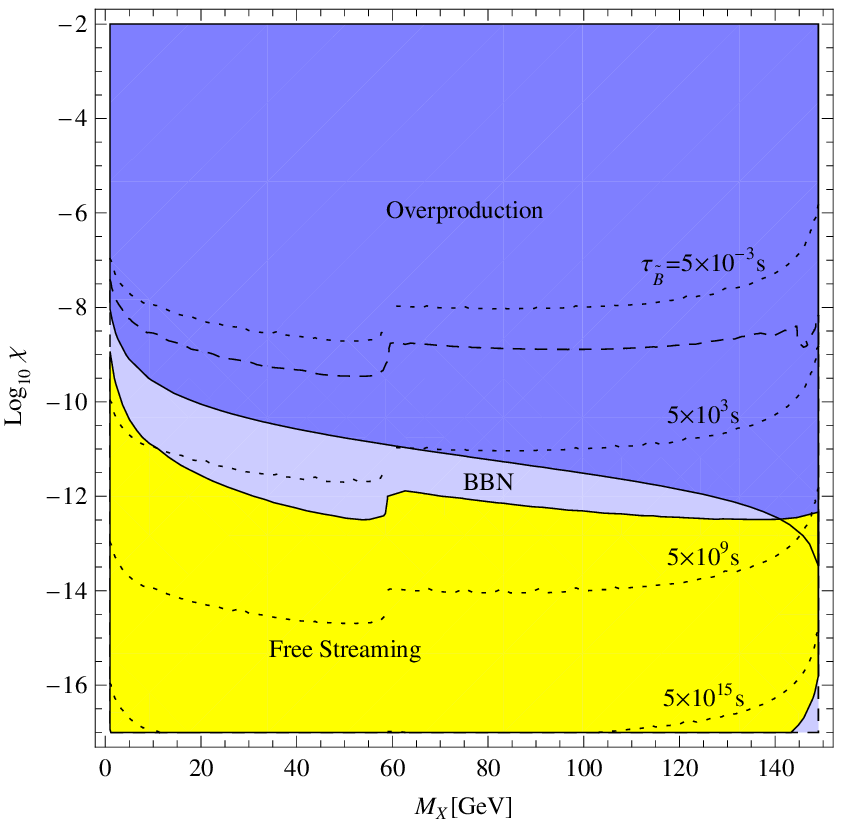}
    \vspace{-0.5cm}
  \end{center}
  \caption{Summary of bounds on the hidden $U(1)$ gaugino parameter space for
  the case of a bino-like neutralino NLSP.  We take $M_B=150\GeV$. The dark
  blue region is excluded by thermal overproduction.  Below this region, the
  hidden $U(1)$ gaugino would be dominantly produced via late decaying
  neutralinos. This scenario is totally excluded by BBN \cite{KKM05} (light
  blue region). The bound actually strongly overlaps with the overproduction
  region (dashed line). We also show the region (in yellow) that would be
  solely excluded by free streaming arguments. The dotted lines show the
  lifetime of the neutralino. (We used $M_\text{sf}=400\GeV$ and $\mu=300\GeV$
  for the branching ratios).} 
  \label{fig:boundsBino}
\end{figure}

In Fig.~\ref{fig:boundsSlepton} (for $M_B=180\GeV$) and
Fig.~\ref{fig:boundsBino} (for $M_B=150\GeV$) we show the region in the
$(\chi,M_X)$ parameter space where the hidden $U(1)$ gaugino is overproduced.
Remarkably, a large part of the parameter space suggested by string theory is
excluded by the constraint Eq.~\eqref{eqn:OverProductionBound}.\footnote{A
sizable annihilation rate between hidden $U(1)$ gauginos would allow to
circumvent this bound.  If the mixing angle $\Theta$ is small, this would
require additional light matter states in the hidden sector, which do not exist
in our scenario.}

\section{Bounds from primordial nucleosynthesis and \\ structure formation}
\label{sec:BBN}
In general, dark matter can be produced thermally and/or in the late decay of
some other relic particle $\xi$. The abundance of dark matter today is then
given by $\Omega_{\text{DM}}=\Omega_{\text{DM}}^{\text{th.}}+ (M_{\text{DM}} /
M_\xi) \Omega_\xi^{\text{th.}}$, where $\Omega_\xi^{\text{th.}}$ refers to the
thermal abundance of the particle $\xi$.  Typical candidates for $\xi$, like
the stau or a neutralino, naturally freeze out with an abundance of the right
order of magnitude to account for the observed dark matter abundance, and a
given dark matter candidate can inherit this property as long as its direct
thermal production is negligible and provided the masses of the two particles 
are comparable.

These so-called superWIMP scenarios (super-weakly interacting massive
particles, see Ref.~\cite{FRT03}), where the dark matter superWIMPs are
produced in the late decay of other particles, are subject to strong
constraints due to their potential impact on primordial nucleosynthesis and
structure formation. If the production takes place during or after Big Bang
Nucleosynthesis (BBN), the hadronic and electromagnetic energy released in the
decay can destroy the successful predictions of the standard scenario (see \fex
Ref.~\cite{KKM05, HHK+07}). In cases where the decaying particle is
electrically charged, it could additionally form bound states with $^4$He,
triggering the catalytic production of $^6$Li and $^9$Be \cite{Pospelov07,
PPS08}.  Furthermore, the superWIMP itself is potentially produced with a
momentum that is large enough to wash out density fluctuations on small scales
with impact on structure formation, similar to scenarios with warm dark matter
(for a discussion in the context of superWIMPs see Ref.~\cite{CFR+05}).

Below, we will apply these bounds to the case of a hidden $U(1)$ gaugino
superWIMP, concentrating on the four different scenarios shown in
Tab.~\ref{tab:Spectra}. Following the superWIMP paradigm, the relic abundance
of the neutralino or the slepton is fixed by the requirement that the LSP
constitutes all dark matter today.

\begin{table}
  \centering
  \begin{tabular}{|c|c|c|c|c|}
    \hline
    & $i)$		& $ii)$		& $iii)$	& $iv)$		\\\hline\hline
    NNLSP	& $\cdot$	& $\cdot$	& slepton	& slepton	\\
    NLSP	& slepton	& neutralino	& gravitino	& $\lambda_X$	\\
    LSP	& $\lambda_X$	& $\lambda_X$	& $\lambda_X$	& gravitino	\\
    \hline
  \end{tabular}
  \caption{The different scenarios that we will investigate in this work.}
  \label{tab:Spectra}
\end{table}

\paragraph{Scenario i)} 
After freeze out, the slepton NLSP decays predominantly through the channel
$\tilde{l}\rightarrow\lambda_X l$ into a lepton and the hidden $U(1)$ gaugino
LSP. The corresponding decay width is given by
\begin{eqnarray}
  \Gamma_{\tilde{l}\rightarrow \lambda_X l} \simeq \frac{g'^2}{8\pi}\Theta^2 
  Y_{\tilde{l}}^2 M_{\tilde{l}} \left( 1- \frac{M_X^2}{M_{\tilde{l}}^2} \right)^2\;,
  \label{eqn:DecayWidthSleptonApprox}
\end{eqnarray}
where $Y_{\tilde{l}}$ and $M_{\tilde{l}}$ denote, respectively, the hypercharge
and the mass of the slepton NLSP. 

The decay induces electromagnetic cascades in the plasma, the branching
fraction into electromagnetic energy being close to one,
$B_\text{EM}^{\tilde{l}}\simeq1$. The corresponding energy release is given by
$\epsilon^{\tilde{l}}_{\text{EM}}\approx(M^2_{\tilde{l}}-M_X^2) /(2
M_{\tilde{l}})$ (see Ref.~\cite{FST04} for an exhaustive discussion).

Contributions to the hadronic energy release stem, if kinematically allowed,
from three-body decays into $Z^0$, $W^{\pm}$ and Higgs bosons. The
corresponding branching ratios are all of the order of, or smaller than,
$\alpha'/(16\pi s_\text{w}^2 )\sim 10^{-2}$, whereas the hadronic branching
fractions of the subsequent decay processes are all of the order of one. If
these decay channels are kinematically closed, the dominant contribution to
hadronic energy release comes from the strongly suppressed four-body decay
$\tilde{l}\rightarrow\lambda_X lq\bar{q}$, with a branching ratio of the order
of $10^{-6}$. Hence, the overall hadronic branching fraction is
$B_\text{had}^{\tilde{l}}\sim 10^{-6}$ for large masses of the hidden $U(1)$
gaugino ($M_X\gtrsim M_{\tilde{l}}-M_Z$) and $B_\text{had}^{\tilde{l}}\sim
10^{-2}$ for small masses ($M_X\lesssim M_{\tilde{l}}-M_Z$). Variations in the
corresponding energy release $\epsilon_\text{had}^{\tilde{l}}$ are subdominant
because $B_\text{had}^{\tilde{l}}$ varies by several orders of magnitude, and
we simply take $\epsilon_\text{had}^{\tilde{l}}=(1/3)(M_{\tilde{l}}-M_X)$
\cite{FST04}.

Applying the BBN bounds from Ref.~\cite{KKM05} to this scenario, a considerable
part of the hidden $U(1)$ gaugino parameter space can be excluded (see
Fig.~\ref{fig:boundsSlepton}).\footnote{In Ref.~\cite{KKM05} the bounds are
shown for the cases $B_\text{had}=1$, $10^{-3}$ and zero. We just interpolate
between $B_\text{had}=1$ and $B_\text{had}=10^{-3}$ linearly in
$\log_{10}(B_\text{had})$ and use $B_\text{had}=0$ as a cutoff.} We also show
the region that is excluded by the catalytic enhancement of $^6\text{Li}$ and
$^9\text{Be}$ production. These effects essentially require that the lifetime
of the slepton is smaller than around $2\times10^{3}\s$ \cite{Pospelov07}.

The impact of superWIMP scenarios on structure formation is encoded in the free
streaming length $\lambda_\text{FS}$, defined as the comoving distance traveled
by a particle since it was produced:
\begin{eqnarray}
  \lambda_\text{FS}=\int_0^{z_\text{prod.}} dz \frac{v(z)}{H(z)}\;.
  \label{eqn:DefFS}
\end{eqnarray}
On scales below the free streaming length of dark matter particles, density
fluctuations are washed out before structure formation begins (see \fex
Ref.~\cite{KT90}). The non-observation of this effect in the latest
Lyman-$\alpha$ Forest data from the Sloan Digital Sky Survey \cite{M06}
excludes a free streaming length larger than roughly $\lambda_\text{FS}\gtrsim
0.5\Mpc$.\footnote{See Ref.~\cite{SKB07}, where the power spectrum of
non-thermally produced dark matter is confronted with lower bounds on the mass
of sterile neutrino dark matter \cite{SMM+06, VLH+06}.} The region in the
parameter space of the hidden $U(1)$ gaugino that is excluded by this
requirement is shown in Fig.~\ref{fig:boundsSlepton} and lies completely in the
region already ruled out by BBN.

\paragraph{Scenario ii)}
Two-body decays of a neutralino NLSP lead to final states where the hidden
$U(1)$ gaugino LSP is accompanied either by a $Z^0$ or Higgs boson or a
photon.  However, if the decay into $Z^0$ or Higgs bosons is kinematically
forbidden, neutralinos predominantly decay via $\lambda_B\rightarrow\lambda_X
f\bar{f}$, since the radiative two-body decay into photons is one-loop
suppressed and negligible in most cases. We have summarized all relevant decay
widths in appendix~\ref{apx:decayWidth}, particularized to the case of a
bino-like neutralino. The hadronic branching fraction for the decay of a
bino-like neutralino is of the order of one for hidden $U(1)$ gaugino masses
that allow the decay into $Z^0$ bosons, and can range between $10^{-2}$ and
order one otherwise (depending on the squark masses). For definiteness we choose
$B_\text{had}^{\lambda_B}\sim\mathcal{O}(1)$ for all masses $M_X$, since
smaller values at large $M_X$ will not change our conclusions.  For the
corresponding energy release we simply take $\epsilon_\text{had} = 1/3
(M_B-M_X)$. We apply the BBN bounds of Ref.~\cite{KKM05} and show the
corresponding excluded region in Fig.~\ref{fig:boundsBino}.  Together with the
overproduction bounds, this scenario is excluded in the whole $(\chi, M_X)$
parameter space suggested by string theory, $\chi\sim 10^{-2}-10^{-16}$.

\begin{figure}[h]
  \begin{center}
    \includegraphics{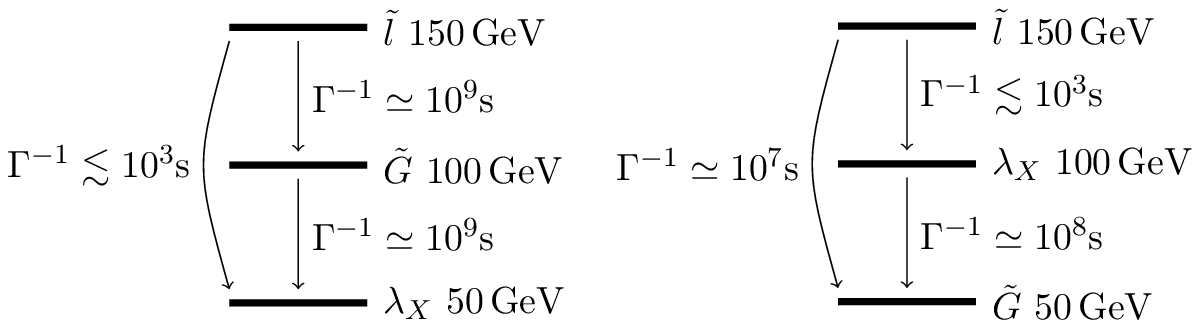}
    \hspace{6cm}a\hspace{6cm}b\hspace{5cm}
  \end{center}
  \caption{Spectra for (a) gravitino NLSP and (b) gravitino LSP (\comp last two
  cases in Tab.~\ref{tab:Spectra}).  We show the widths for the different decay
  processes for typical particle masses.  The mixing parameter is assumed to
  lie in the allowed region of Fig.~\ref{fig:boundsSlepton}.}
  \label{fig:SpectrumWithGravitinos}
\end{figure}

\begin{figure}[h]
  \begin{center}
    \includegraphics[width=0.7\linewidth]{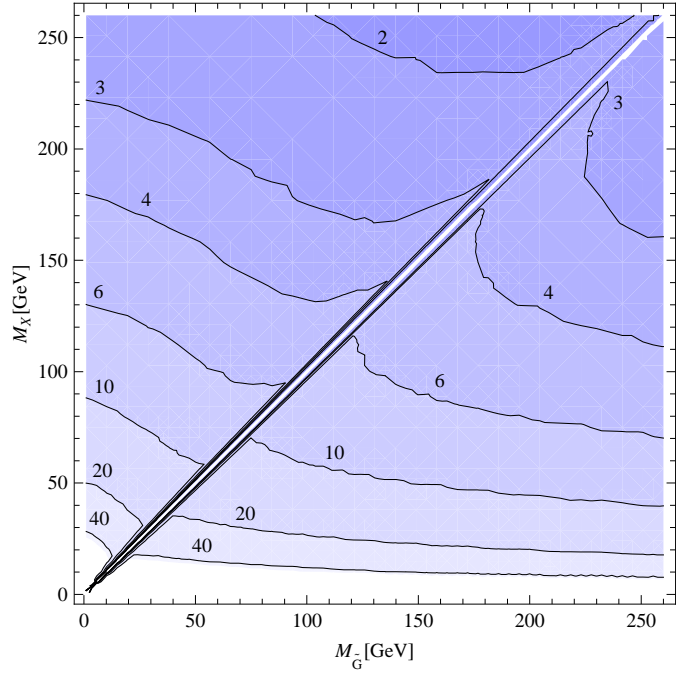}
  \end{center}
  \caption{Contour plot of free streaming lengths in units of $\text{Mpc}$. The
  upper left (lower right) corner shows the free streaming length of gravitinos
  (hidden $U(1)$ gauginos) that stem from the late decay of hidden $U(1)$
  gauginos (gravitinos).}
  \label{fig:FSXandGravitino}
\end{figure}

\begin{figure}[h]
  \begin{center}
    \includegraphics[width=0.7\linewidth]{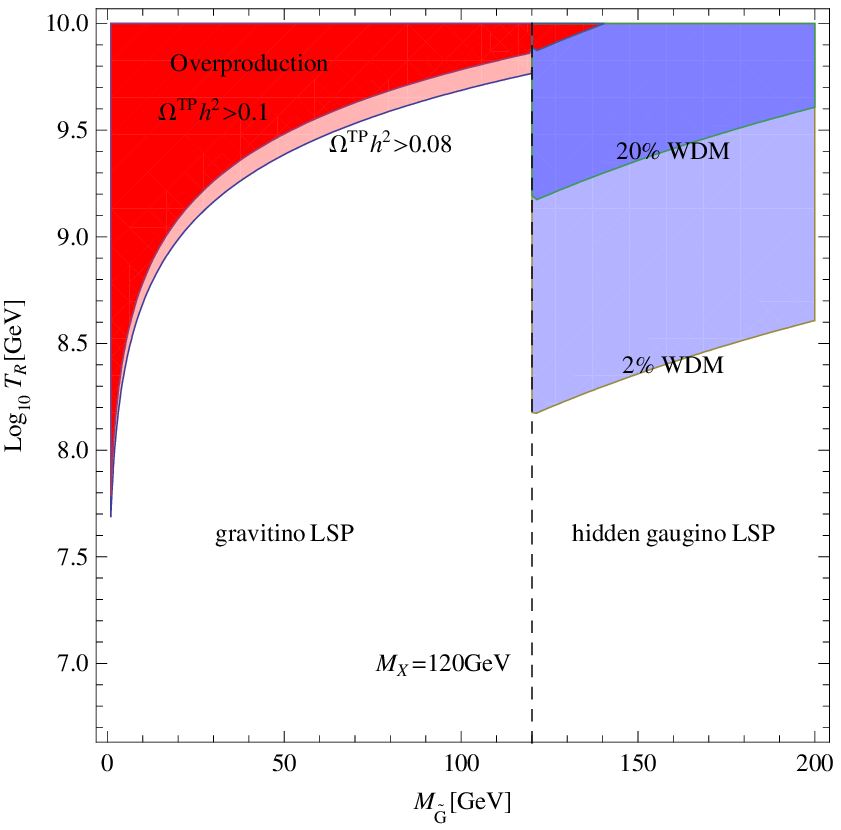}
  \end{center}
  \caption{Bounds on the reheating temperature as function of the gravitino
  mass, using Eq.~\eqref{eqn:GravitinoProduction} with $m_{\tilde{g}}=800\GeV$.
  The mass of the hidden $U(1)$ gaugino is fixed to $M_X=120\GeV$.  If the
  gravitino is the LSP (left part), the reheating temperature is only bounded
  by overproduction arguments (dark red region), which are only slightly
  strengthened when, say, 20\% of the gravitino abundance is due to non-thermal
  production (light red region). A gravitino NLSP (right part) would late decay
  into the hidden $U(1)$ gaugino, yielding a warm dark matter component. If
  only a fraction of $20\%$ or $2\%$ of dark matter is allowed to be warm (with
  free streaming lengths as shown in Fig.~\ref{fig:FSXandGravitino}), the
  corresponding blue regions are excluded.}
  \label{fig:FSGravitinoNLSP}
\end{figure}

\paragraph{Scenario iii)}
We will now discuss the effects of an additional light gravitino, concentrating
on the spectra shown in Fig.~\ref{fig:SpectrumWithGravitinos}, where we also
show typical values for the relevant decay widths. 

If the hidden $U(1)$ gaugino is the LSP (see
Fig.~\ref{fig:SpectrumWithGravitinos}a), it can be produced in three different
ways: thermally, via slepton decay and via gravitino decay.  For mixings and
masses that are allowed in Fig.~\ref{fig:boundsSlepton}, the decay of sleptons
into gravitinos is strongly suppressed, and the decay of gravitinos into hidden
$U(1)$ gauginos is only accompanied by hidden $U(1)$ gauge bosons and invisible
to MSSM particles.  Hence, bounds from BBN and overproduction are essentially
the same than in case $i)$.

However, this does not hold for the free streaming bounds, since, in contrast
to hidden $U(1)$ gauginos which are produced thermally or via slepton decay,
hidden $U(1)$ gauginos that stem from the decay of gravitinos can have a quite
large free streaming length, of the order of several $\text{Mpc}$, as shown in
Fig.~\ref{fig:FSXandGravitino}. This leads to dark matter with a cold and a
warm component, so-called $\Lambda$CWDM models.

Constraints on $\Lambda$CWDM models where recently studied in
Refs.~\cite{VLH+05, PCS+07, IMM08}. A warm dark matter component would induce a
step in the power spectrum of density fluctuations, with a size which depends
on the fraction of the dark matter that is warm and a position which is roughly
given by the corresponding free streaming length as $k_\text{FS}\sim
2\pi/\lambda_\text{FS}$. Although models with mixed cold/warm dark matter arise
quite naturally in many situations, a full general analysis with all the latest
data is still lacking.\footnote{In Ref.~\cite{IMM08} a consideration of the
uncertainty of the power spectrum from WMAP 1-yr data \cite{S03} bounds the
fraction $f$ of dark matter that is allowed to have $\lambda_\text{FS} \simeq
6\Mpc$ to be $f\lesssim 0.2 - 0.4$. This seems to be consistent with a
hydrodynamical analysis in Ref.~\cite{VLH+05}, where the bound on the fraction
of dark matter that can be made out of thermal light gravitinos with a mass of
$ m_{\tilde{G}} \approx 16\eV$ is stated as $f\lesssim 0.12$, since the
corresponding free streaming length (using Eq.~\eqref{eqn:DefFS}) is around
$\lambda_\text{FS}\sim 40\Mpc$.  However, both analyses do not take into
account the latest Lyman-$\alpha$-forest data \cite{M06}. An analysis in
Ref.~\cite{PCS+07} which incorporates these data, and where lower bounds on the
mass of sterile neutrino dark matter from Ref.~\cite{SMM+06} are rescaled for
the case of $\Lambda$CWDM models, suggests that even a component with a free
streaming lengths around $\lambda_\text{FS}\sim 4\Mpc$ is bounded strongly with
$f\lesssim 0.1$.} Bearing in mind this limitation, we will simply show the
corresponding bounds on the reheating temperature and mixing parameter for the
two reference values $f=0.2$ and $f=0.02$, where $f$ denotes the fraction of
dark matter that is warm, $\Omega_\text{DM}=f
\Omega_\text{WDM}+(1-f)\Omega_\text{CDM}$; a complete analysis of the mixed
cold/warm dark matter scenario is beyond the scope of this paper.

As discussed above, the warm component of hidden $U(1)$ gaugino DM comes from
the late decay of gravitinos. The thermal abundance of gravitinos directly
depends on the reheating temperature $T_R$ like \cite{BBB01,PS07,RS07}
\begin{eqnarray}
  \Omega_{\tilde{G}}h^2\simeq0.27 \left( \frac{T_R}{10^{10}\GeV} \right) 
  \left( \frac{100\GeV}{m_{\tilde{G}}} \right) \left( 
  \frac{m_{\tilde{g}}}{1\text{TeV}} \right)^2\;,
  \label{eqn:GravitinoProduction}
\end{eqnarray}
where $m_{\tilde{g}}$ denotes the gluino mass. Hence, the bounds on
$\Lambda$CWDM models translate into a bound on the reheating temperature as
shown in the right part of Fig.~\ref{fig:FSGravitinoNLSP}. 

\paragraph{Scenario iv)}
Scenarios with a gravitino LSP and a slepton (usually a stau) NLSP are known to
be strongly constrained due to the catalytic production of $^6\text{Li}$ and
$^9\text{Be}$ in the presence of long-lived charged particles during BBN (see
above discussion). The resulting bound on the lifetime of staus translates into
an upper bound on the gravitino mass of around $m_{\tilde{G}}\lesssim 1\GeV$.
At the same time, requiring that gravitinos with $m_{\tilde{G}}\lesssim 1\GeV$
are not overproduced forbids reheating temperatures as high as
$T_R\sim10^9\GeV$ (see left part of Fig.~\ref{fig:FSGravitinoNLSP} and
Eq.~\eqref{eqn:GravitinoProduction}). However, these high reheating temperatures
are favored in scenarios with leptogenesis as the source of the baryon
asymmetry of the universe \cite{FY86, BDP05}. 

One way to relax the tension between leptogenesis and gravitino DM is to assume
that the production of entropy between stau freeze out and BBN dilutes the stau
abundance sufficiently to evade the bounds \cite{BHI+06, PS07a}. Another way is
to impose a small violation of $R$-parity \cite{BCH+07}, which allows the stau
to decay fast into standard model particles. In this case, the gravitino
becomes unstable and decays with a lifetime which is typically longer than the
age of the universe. The gravitino decay products could be detected as an
anomalous contribution to the cosmic ray fluxes \cite{IT08, IT08a, IMM08a},
opening the possibility to probe this scenario.

Interestingly, a similar mechanism can work in the presence of a hidden $U(1)$
gaugino for spectra of the form (slepton, $\lambda_X$, gravitino), as depicted
in Fig.~\ref{fig:SpectrumWithGravitinos}b. For mixings and masses of the hidden
$U(1)$ gaugino that lie in the allowed region of Fig.~\ref{fig:boundsSlepton},
the lifetime of the stau is compatible with all BBN bounds. In this case, the
reheating temperature can be as large as $T_R\sim10^9\GeV$, as shown in
Fig.~\ref{fig:FSGravitinoNLSP}, since the gravitino mass and the stau lifetime
are decoupled.

The warm component of gravitino dark matter, with a free streaming length of
the order of several $\text{Mpc}$ (see Fig.~\ref{fig:FSXandGravitino}), comes
now from the late decay of hidden $U(1)$ gauginos. Hence, bounds on
$\Lambda$CWDM models constrain the sum over the thermal and non-thermal
abundance of hidden $U(1)$ gauginos in this case. Firstly, this shrinks the
allowed region of the hidden $U(1)$ gaugino parameter space as shown in
Fig.~\ref{fig:boundsSlepton} (dashed lines). Secondly, in cases where the
thermal production of hidden $U(1)$ gauginos is negligible, the stau abundance
is bounded like $Y_\text{stau} \lesssim (9\times10^{-14} - 9\times10^{-13})
(M_{\tilde{G}}/100\GeV)^{-1}$ if we allow a fraction $f=0.02 - 0.2$ of dark
matter to be warm. Thus, the $^9$Be and $^6$Li bounds on the abundance of
long-lived staus, which are roughly given by $Y_\text{stau}\lesssim 10^{-15}$
and $Y_\text{stau}\lesssim 10^{-16}$, respectively (see Ref.~\cite{PPS08}), can
be weakened by two or three orders of magnitude. 

In the CMSSM with gravitino dark matter the typical relic abundance of stau
NLSPs ranges between $Y_\text{stau} \sim10^{-14}$ and $Y_\text{stau}
\sim10^{-13}$ (although considerably smaller values are possible due to
resonant effects or coannihilation \cite{PS08}, or \fex in models with
non-universal Higgs masses \cite{RSW08}). In these kind of models, the
phenomenologically allowed parameter space will be greatly enlarged by the
above mechanism. However, a detailed analysis of the influence of a weakly
interacting hidden sector on constraints on the reheating temperature would
require reliable bounds for general $\Lambda$CWDM models.

\section{Collider phenomenology}
At the LHC, cascade decays of squarks and gluinos can produce of the order of
$10^6$ NLSPs per year if the sparticle masses are close to the current
experimental limits~\cite{BHS+97}, thus offering a unique opportunity to test
scenarios with a light hidden $U(1)$ gaugino. It was shown in the previous
sections that in this class of scenarios the neutralino as NLSP is excluded,
whereas the stau as NLSP is allowed only if it has a lifetime that ranges
between $\tau\sim 10^{-2}\s$ and $\tau\sim 10^{3}\s$. Therefore, the stau will
typically traverse the detector leaving a distinctive heavily ionizing charged
track~\cite{DT90}. Moreover, if the stau is slow enough, it might get trapped
in the detector, decaying at late times and thus allowing a measurement of the
lifetime and, correspondingly, of the mixing parameter $\chi$. The number of
staus trapped can be incremented by placing a 1-10 kton massive material around
the LHC detectors, which will allow to collect ${\cal O}(10^3-10^4)$
staus~\cite{HKN+04}. If the stopper material is simultaneously an active
real-time detector, the decay products and their angular distribution could be
studied in detail. This will allow to distinguish this scenario from the
scenario of gravitino dark matter, which yields very similar signatures at
colliders~\cite{BHR+04}.

\section{The case with an anomalously small $\chi$}
We have shown in sections~\ref{sec:OP} and \ref{sec:BBN} that constraints from
overproduction, BBN and structure formation exclude the scenario with a
neutralino NLSP and hidden $U(1)$ gaugino for natural values of the mixing
parameter $\chi\sim 10^{-2}-10^{-16}$ (see Fig.~\ref{fig:boundsBino}).
However, these strong constraints could be evaded in more elaborated models
yielding a smaller $\chi$. Our interest in exploring the region with small
$\chi$ is that the neutralino lifetime could become larger than the age of the
Universe, thus constituting a viable dark matter candidate
itself.\footnote{Clearly, the scenario with charged slepton NLSP is excluded
for very small values of $\chi$, since it would yield an abundance of
anomalously heavy hydrogen in conflict with the experimental constraints.}
Nevertheless, in contrast to the standard neutralino dark matter scenario, in
this scenario the neutralino NLSP is unstable and decays with very long
lifetimes into the hidden gaugino and Standard Model particles.

The neutralino dark matter particle (that for definiteness we assume to be a
pure bino) can decay into different channels:
\begin{eqnarray}
  \lambda_B& \rightarrow& Z\; \lambda_X\;, \nonumber \\
  \lambda_B& \rightarrow& \gamma\; \lambda_X\;, \nonumber \\
  \lambda_B&\rightarrow& h\; \lambda_X\;, \nonumber \\
  \lambda_B& \rightarrow& f \bar f\; \lambda_X\;,
\end{eqnarray}
with decay rates listed in appendix~\ref{app:DR}. The fragmentation of the
$Z^0$ and the Higgs boson produces a continuous spectrum of stable particles,
such as photons, positrons or antiprotons which contribute to the primary
fluxes of cosmic rays. The experimental constraints on these fluxes will then
translate into constraints on the parameters of the model.

On the other hand, it has been pointed out recently that the decay of a dark
matter particle with a lifetime $\tau\sim 10^{26}$s into weak gauge bosons with
a momentum $\sim 50$ GeV could simultaneously explain the EGRET anomaly in the
extragalactic gamma-ray background and the HEAT excess in the positron
fraction \cite{IT08a, IMM08a}. Interestingly, the scenario we are 
considering here, where the neutralino dark matter particle decays into a $Z^0$
gauge boson, constitutes a promising candidate of this kind.  Moreover, in
contrast to other candidates for decaying dark matter that have been proposed,
such as the gravitino \cite{BCH+07} or a hidden $U(1)$ gauge boson
\cite{CTY08}, the unstable neutralino could be detected in direct dark matter
searches.

Although the simplest models with hidden gaugino LSP and neutralino NLSP
predict a mixing parameter $\chi$ ranging between $10^{-2}$ and $10^{-16}$, and
thus a neutralino lifetime ranging between $\tau\sim10^{-14}\s$ and
$\tau\sim10^{14}\s$, it is not difficult to construct plausible models where
the neutralino lifetime can be orders of magnitude longer than the age of the
Universe.  To this end, let us consider a model with two hidden $U(1)$ gauge
groups, $U(1)_{X}$ and  $U(1)_{X^\prime}$, and vector superfields denoted by
$\hat X$ and $\hat X^\prime$, respectively.  We will assume that the hidden
sector particles that generate the kinetic mixing are charged either under
$U(1)_Y$ and $U(1)_{X^\prime}$, or under $U(1)_X$ and $U(1)_{X^\prime}$, but
not under $U(1)_Y$ and $U(1)_{X}$ simultaneously. If this is the case, the
kinetic mixings $\hat{W}_B^\alpha\hat{W}_{X^\prime\,\alpha}$ and
$\hat{W}_{X}^\alpha\hat{W}_{X^\prime\alpha}$ will be generated, but not
$\hat{W}_B^\alpha\hat{W}_{X\,\alpha}$. Therefore, the SUSY invariant part of
the Lagrangian reads
\begin{eqnarray}
  \label{eqn:twohiddenU1sSUSY}
  {\cal L}_{\rm SUSY} &=&\int d^2\theta \, \left(\hat{W}_B^\alpha \hat{W}_{B\,\alpha}+
  \hat{W}_{X}^\alpha \hat{W}_{X\,\alpha}+
  \hat{W}_{X^\prime}^\alpha \hat{W}_{X^\prime\,\alpha}+\right. \nonumber \\
  && \left.2\chi_1 \hat{W}_B^\alpha\hat{W}_{X^\prime\,\alpha}+
  2\chi_2 \hat{W}_{X}^\alpha\hat{W}_{X^\prime\alpha}\right)
  +{\rm h.c.} \;.
\end{eqnarray}
In addition, there exists a soft SUSY breaking Lagrangian:
\begin{equation}
  {\cal L}_{{\rm SUSY}\hspace{-0.75cm}\line(4,1){20}}=-\frac{1}{2}
  \ppp{\hat{\lambda}_X &\hat{\lambda}_{X^\prime} & \hat{\lambda}_B}\hat{\mathcal{M}}
  \ppp{\hat{\lambda}_{X}\\\hat{\lambda}_{X^\prime}\\\hat{\lambda}_{B}} + 
  \text{h.c.}\;,
  \label{eq:twohiddenU1ssoft}
\end{equation}
where 
\begin{equation}
  \hat{\mathcal{M}}=\ppp{\hat{M}_X&\delta\hat{M_2}& 0\\
  \delta\hat{M_2}&\hat{M}_{X^\prime}&\delta\hat{M_1}\\
  0&\delta\hat{M_1}&\hat{M_B}}\;.
\end{equation}
This structure for the soft mass matrix can be straightforwardly derived from
Eq.~\eqref{eqn:twohiddenU1sSUSY} introducing five arbitrary spurion superfields
as in Eq.~\eqref{eqn:LagrangianSuperFields}.  Note that $\delta\hat{M_1}$ and
$\delta\hat{M_2}$ are suppressed by $\chi_1$ and $\chi_2$, respectively.

In order to canonically normalize the vector superfields, we define
\begin{eqnarray}
  B&=&\hat B \;, \nonumber \\
  X&=&\hat X-\chi_1\, \chi_2 \hat B  \;,\nonumber \\
  X^\prime &=& \hat X^\prime +\chi_1 \hat B+\chi_2 \hat X \;,
\end{eqnarray}
where we have assumed  $\chi_1,\chi_2\ll 1$. Then, the SUSY invariant part of
the Lagrangian reads:
\begin{eqnarray}
  {\cal L}_{\rm SUSY}&=&\int d^2\theta \, \left(W_B^\alpha W_{B\,\alpha}+
  W_{X}^\alpha W_{X\,\alpha}+
  W_{X^\prime}^\alpha W_{X^\prime\,\alpha}\right)
  +{\rm h.c.} \;,
\end{eqnarray}
while the SUSY breaking Lagrangian reads:
\begin{equation}
  {\cal L}_{{\rm SUSY}\hspace{-0.75cm}\line(4,1){20}}=-\frac{1}{2}
  \ppp{\lambda_X &\lambda_{X^\prime} & \lambda_B}\mathcal{M}
  \ppp{\lambda_{X}\\ \lambda_{X^\prime}\\\lambda_{B}} + {\rm h.c.}\;,
\end{equation}
with
\begin{equation}
  \mathcal{M}\simeq \ppp{ M_X&\delta M_2& \delta M_{12}\\
  \delta M_2&M_{X^\prime}&\delta M_1\\
  \delta M_{12}&\delta M_1&M_B}\;.
\end{equation}
In this equation, $M_X\simeq \hat M_X$, $M_{X^\prime}\simeq \hat M_{X^\prime}$
and $M_B\simeq \hat M_B$, while $\delta M_2\simeq \delta\hat{M_2}-\chi_2\,
M_{X^\prime}$ and $\delta M_1\simeq \delta\hat{M_1}-\chi_1\, M_{X}$. More
importantly, after the canonical normalization of the kinetic terms, a mixing
term between the bino and the hidden gaugino $\lambda_X$ has been generated,
$\delta M_{12}\simeq \chi_1\, \chi_2\,(M_X+M_{X^\prime}) - \chi_2\,
\delta\hat{M_1} - \chi_1\,\delta\hat{M_2}$, which is doubly suppressed by
$\chi_1$ and $\chi_2$. 

If $\lambda_X$ is the LSP and $\lambda_B$ is the NLSP, the decay rate of the
bino NLSP is very strongly suppressed, thus yielding very long lifetimes for
the neutralino dark matter. For instance, assuming that the kinetic mixing
between the observable sector and the hidden sector is $\chi_1\sim 10^{-16}$,
as suggested by string theory, a mixing between the two hidden sector $U(1)$s
of $\chi_2 \sim 10^{-7}$ would be necessary in order to render a neutralino
lifetime of the order of $10^{26}\s$.

The signatures in cosmic ray experiments of the scenario of neutralino dark
matter which decays into hidden gauginos are fairly model dependent and will be
shown elsewhere \cite{prep}. In this paper we will just derive 
a conservative bound on the mixing parameter $\chi$ from requiring a gamma ray
flux from neutralino decay in agreement with the EGRET observations. To this
end, we will just consider the contribution to the gamma ray flux from the two
body decay $\lambda_B\rightarrow \gamma\, \lambda_X$. 

The total gamma ray flux receives several contributions.  First, we expect a
gamma ray flux from the center of the Galaxy produced by neutralino
annihilations.  This contribution does not depend on whether the neutralino is
absolutely stable or on the contrary decays at late times.  Hence, it does not
constrain our scenario and will not be further discussed. The interested reader
can find in the vast existing literature the prospects of detecting such a
signal from neutralino annihilation coming from the center of the Galaxy
(see \fex Ref.~\cite{BHS05}).

\begin{figure}[h]
  \begin{center}
    \includegraphics[width=0.9\linewidth]{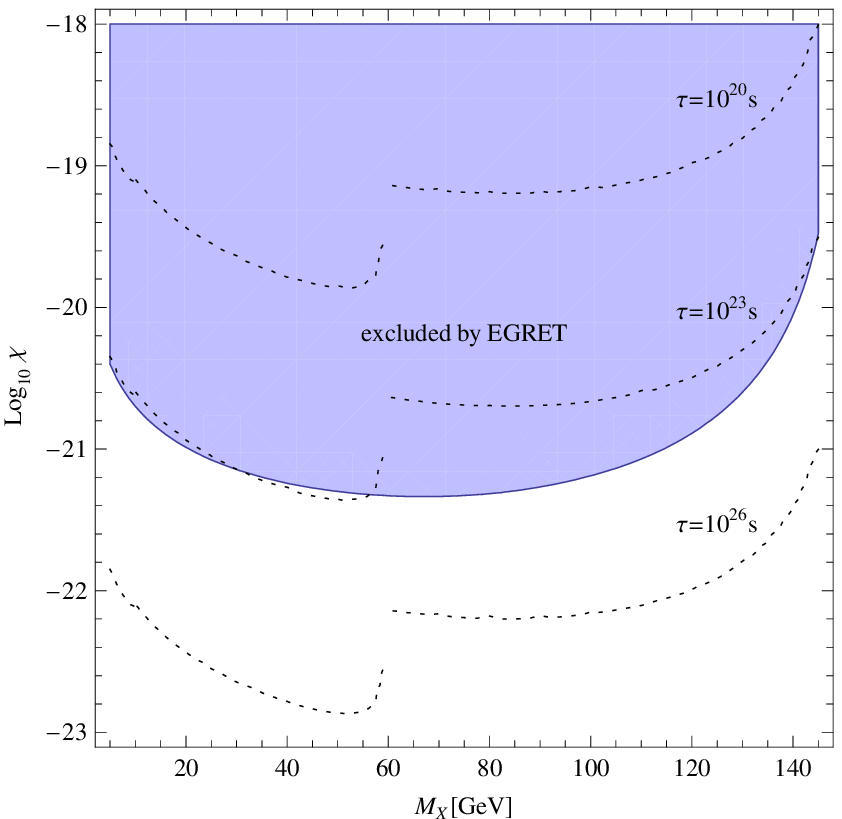}
  \end{center}
  \caption{Region of the hidden $U(1)$ gaugino parameter space which is
  excluded by confronting the EGRET measurements \cite{S98} with the 
  extragalactic gamma ray flux produced in the decay 
  $\lambda_B\rightarrow\gamma\lambda_X$ (blue region). 
  The bino mass is fixed to $M_B=150\GeV$. We also show
  for reference the isocurves for the neutralino life-time.}
  \label{fig:EGRETbounds}
\end{figure}

In addition, we expect a diffuse flux produced in the decay of neutralinos at
cosmological distances and in the Milky Way halo. Following Ref.~\cite{BCH+07,
BBC+07}, we will average the halo gamma ray signal of the $\lambda_B\rightarrow
\gamma \lambda_X$ channel over the whole sky, excluding a band of $\pm
10^\circ$ around the Galactic disk in order to compare our results with the
EGRET data~\cite{SMR04}.  To be conservative, we will require that the
corresponding peak in the energy spectrum, after convolving with the energy
resolution of 15\% quoted by EGRET, remains below the 2$\sigma$ band of the
EGRET spectrum obtained by Sreekumar {\it et al.} in~\cite{S98}, 
\begin{equation}\label{sreekpowerlaw}
  \frac{dJ_{\rm EGRET}}{dE} =
  (7.32 \pm 0.34) \times 10^{-6}  
  ({\rm cm}^2~{\rm s}~{\rm str}\  {\rm GeV})^{-1} 
  \left(\frac{E}{0.451 \ {\rm GeV}}\right)^{-2.1\pm 0.03}\; .
\end{equation}
The result is shown in Fig.~\ref{fig:EGRETbounds}, where we used a reference
bino mass of $M_B=150\GeV$. For hidden $U(1)$ gaugino masses around
$M_X\sim75\GeV$, mixings down to $\chi\sim 10^{-21}$ can be excluded 
from the EGRET measurements of the gamma ray flux. 

\section{Conclusions}
An unbroken hidden $U(1)$ that interacts with the Standard Model only via
kinetic mixing with hypercharge decouples completely from the visible world.
However, in the supersymmetric version of this scenario this is no longer the
case. We have shown that a mass mixing between the bino and the hidden $U(1)$
gaugino is always generated via radiative effects, although this mixing can be
generated  already at tree level in some well motivated scenarios.  Moreover,
we have discussed different scenarios of supersymmetry breaking in which the
hidden $U(1)$ gaugino is the lightest supersymmetric particle.

We have mostly concentrated on this possibility and we have derived
cosmological bounds on this scenario from precluding overproduction of hidden
$U(1)$ gauginos and from the requirements of successful Big Bang
nucleosynthesis and structure formation.  The combination of these constraints
excludes a neutralino NLSP except for extremely small mixings (see
Fig.~\ref{fig:boundsBino}). On the other hand, when the NLSP is a stau, an
allowed window for mixings around $\chi\sim 10^{-10} - 10^{-13}$ remains (see
Fig.~\ref{fig:boundsSlepton}). In this window, the stau has a lifetime larger
than $10^{-1}\s$ and thus might be detected at future colliders as a heavily
ionizing charged track.

The reheating temperature in scenarios with gravitino dark matter and stau
NLSPs is known to be strongly constrained, and we have shown that these
constraints relax considerably in the presence of a hidden $U(1)$ gaugino (see
Fig.~\ref{fig:FSGravitinoNLSP}).  This might be a rather general effect of very
weakly interacting hidden sectors and deserves further attention.

Finally, we have discussed the case of an anomalously small mixing parameter,
$\chi\ll 10^{-16}$. For these small mixings, the neutralino NLSP can become
long lived enough to constitute the dark matter of the Universe.  We have
constructed a simple model with two hidden $U(1)$s where a tiny mixing can be
naturally obtained. Even though the neutralino is very long lived, it
eventually decays into the hidden gaugino and standard model particles, which
might be detected as an anomalous contribution to the cosmic ray fluxes.  Using
the EGRET measurement of the extragalactic gamma ray flux, we have derived a
conservative bound on the mixing parameter $\chi\lesssim 10^{-20-21}$ (see
Fig.~\ref{fig:EGRETbounds}).

\section*{Acknowledgments}
 C.W. thanks Chlo\'e Papineau, James Wells, and especially Javier Redondo for
 valuable discussions.

\begin{appendix}
  \section{Decay rates}
  \label{app:DR}
  For convenience, we summarize the neutralino, slepton and gravitino decay
  rates that we used in this paper.

  The decay widths for bino-like neutralinos that decay into hidden $U(1)$
  gauginos can be derived from the results in Ref.~\cite{HW89, BFM86}. They are
  \label{apx:decayWidth}
  \begin{eqnarray}
    \Gamma(\lambda_B\rightarrow Z^0 \lambda_X)&\simeq&
    \frac{1}{128\pi}g'^2 s_W^2 \Theta^2 M_B
    \lambda^{1/2}\left( 1, \frac{M_X^2}{M_B^2},\frac{M_Z^2}{M_B^2}\right)\times 
    \nonumber\\&&
    \hspace{-3.0cm}\times
    \frac{M_Z^6}{\mu^4 M_B^2}
    \left( \frac{M_B^2}{M_Z^2} + \frac{M_X^2}{M_Z^2}-2
    +\left( \frac{M_B^2}{M_Z^2} - \frac{M_X^2}{M_Z^2}\right)^2 +
    6 \frac{M_X}{M_Z}\frac{M_B}{M_Z}\right)\;,
    \label{eqn:GammaBinoToZX}\\
    %**********************************************************
    \Gamma(\lambda_B\rightarrow \gamma \lambda_X)&\simeq&
    \frac{1}{128\pi}\left( \frac{15}{32\pi^2} \right)^2 e^2 g'^4 \Theta^2
    M_B\times \nonumber\\
    &&\times 
    \frac{M_B^4}{M_\text{sf}^4} \left( 1-\frac{M_X^2}{M_B^2} \right)^3 \left( 
    1-\frac{M_X}{M_B} \right)^2\;,
    \label{eqn:GammaBinoTogammaX}\\
    %**********************************************************
    \Gamma(\lambda_B\rightarrow h \lambda_X)&\simeq&
    \frac{1}{32\pi} g'^2 \Theta^2 M_B 
    \frac{M_Z^2 s_W^2}{\mu^2} 
    \lambda^{1/2}\left( 1, \frac{M_X^2}{M_B^2},\frac{M_h^2}{M_B^2}\right)\times 
    \nonumber\\&&\times
    \left( 1+\frac{M_X^2}{M_B^2} - \frac{M_h^2}{M_B^2} 
    +2\frac{M_X}{M_B} \right)\;,
    \label{eqn:GammaBinoToHiggsX}\\
    %**********************************************************
    \Gamma(\lambda_B\rightarrow f\bar{f} \lambda_X)&\simeq&
    %\frac{9}{2048\pi^3} 
    1.4\times10^{-4} g'^4 \Theta^2 M_B 
    \frac{M_B^4}{M_\text{sf}^4}
    \left(1-\frac{M_X^2}{M_B^2}\right)^5\;,
    \label{eqn:GammaBinoToLeptons}
  \end{eqnarray}
  where we have used the function
  \begin{eqnarray}
    \lambda(a^2,b^2,c^2)=(a^2-(b+c)^2)(a^2-(b-c)^2)\;.
    \nonumber
  \end{eqnarray}
  The approximations that we use for the neutralino mixing angles become exact
  in the limit $M_X \ll M_B \ll M_W \ll \mu$.  Note that the two-body decay
  into a hidden $U(1)$ gaugino and a photon, Eq.~\eqref{eqn:GammaBinoTogammaX},
  is one-loop suppressed, and we sum over all (s)leptons in the loop, assuming
  that they have a common mass $M_\text{sf}$.  The final state in
  Eq.~\eqref{eqn:GammaBinoToLeptons} incorporates neutrinos and charged
  leptons, and we took into account sfermions and $Z^0$ gauge bosons in the
  intermediate state, where the later are subdominant.

  The relevant decay widths with gravitinos in the initial or final state can
  be found in Ref.~\cite{FST04,KMY06} and are
  \begin{eqnarray}
    \Gamma(\lambda_X\rightarrow \tilde{G}X) &=& \frac{1}{48\pi M_\text{pl}^2}
    \frac{M_X^5}{M_{\tilde{G}}^2}
    \left( 1-\frac{M_{\tilde{G}}^2}{M_X^2} \right)^3
    \left( 1+3\frac{M_{\tilde{G}}^2}{M_X^2} \right)\;, \\
    %***
    \Gamma(\tilde{G}\rightarrow \lambda_X X) &=& \frac{1}{32\pi M_\text{pl}^2}
    M_{\tilde{G}}^3
    \left( 1-\frac{M_X^2}{M_{\tilde{G}}^2} \right)^3
    \left( 1+\frac{1}{3}\frac{M_X^2}{M_{\tilde{G}}^2} \right)\;,\\
    %***
    \Gamma(\tilde{l}\rightarrow \tilde{G}l) &=& \frac{1}{48\pi M_\text{pl}^2}
    \frac{M_{\tilde{l}}^5}{M_{\tilde{G}}^2}
    \left( 1-\frac{M_{\tilde{G}}^2}{M_{\tilde{l}}^2} \right)^4\;.
  \end{eqnarray}

  \section{Renormalization group equations}
  \label{app:RG}
  The RGEs for multiple $U(1)$ models with kinetic mixing where first published
  in Ref.~\cite{ACQ88}. One-loop RGEs for gaugino masses and their mixings can
  be found in Ref.~\cite{Suematsu99}.\footnote{Note that our result for the
  RGEs of the gaugino mass matrix in Eq.~\eqref{eqn:mRGplain} differs from the
  one given in Ref.~\cite{Suematsu99} by an additional symmetrization.} 

  The RGEs acquire their simplest form in the basis where gauge bosons and
  gauginos have a canonical kinetic term. The matrix of the gauge couplings
  $\bar{g}_{ij}$ is defined according to the term $\mathcal{L}\supset
  \bar{g}_{ij}j^{\mu}_i A_\mu^j$ in the Lagrangian, where the indices $i=1,2$
  and $j=1,2$ run over the charged currents of the two sectors and over the
  gauge boson states, respectively.  In this representation, the one-loop RGEs
  for the coupling constants are
  \begin{eqnarray}
    \frac{d}{dt}\bar{g}=\frac{1}{16\pi^2}\bar{g}\bar{g}^T B \bar{g}\;,
    \label{eqn:gRGplain}
  \end{eqnarray}
  where we have used the charge matrix $B_{ij}=tr(Q_i Q_j)$, and $t=\ln(Q/Q_0)$
  with $Q$ as RG scale.  The trace in $B_{ij}$ runs over all chiral
  supermultiplets. 

  The RGEs for the gaugino mass matrix $\bar{\mathcal{M}}$ take a similar form
  \begin{eqnarray}
    \frac{d}{dt}\bar{\mathcal{M}}=\frac{1}{16\pi^2}
    \left( \bar{\mathcal{M}}\bar{g}^T B \bar{g}+\bar{g}^T B \bar{g}
    \bar{\mathcal{M}} \right)\;.
    \label{eqn:mRGplain}
  \end{eqnarray}

  Note that only three of the four entries of $\bar{g}$ are physical because
  the basis of the gauge bosons is only fixed up to a rotation. It is
  convenient to state the RGEs for the case where the non-diagonal elements in
  the couplings $g_{ij}$ are chosen to vanish, and where their third free
  component is absorbed in the kinetic mixing term like in
  Eq.~\eqref{eqn:Lagrangian}. We obtain
  \begin{eqnarray}
    \frac{d}{dt}g_{X,B}=\frac{1}{16\pi^2}g_{X,B}^3 B_{XX,BB}\;,
    \label{eqn:gRGint}
  \end{eqnarray}
  \begin{eqnarray}
    \frac{d}{dt}\chi=-\frac{1}{8\pi^2}g_X g_B B_{XB}
    +\frac{1}{16\pi^2}\chi\left( g_X^2 B_{XX} + g_B^2 B_{BB} \right)
    +\mathcal{O}(\chi^2)\;,
    \label{eqn:chiRGint}
  \end{eqnarray}
  \begin{eqnarray}
    \frac{d}{dt}\hat{M}_{X,B}=\frac{1}{8\pi^2}g_{X,B}^2 B_{XX,BB} \hat{M}_{X,B}\;,
    \label{eqn:mRGint}
  \end{eqnarray}
  \begin{eqnarray}
    \frac{d}{dt}\delta\hat{M}= \frac{1}{16\pi^2}\left( g_X^2 B_{XX} + g_B^2 B_{BB} 
    \right)\delta\hat{M}+\mathcal{O}(\chi^2)\;.
    \label{eqn:mmixRGint}
  \end{eqnarray}
\end{appendix}
\frenchspacing
\bibliographystyle{h-physrev3}
\bibliography{hp.bib}
\end{document}